\documentstyle[amssymb,11pt]{article}

\input tcilatex
\if@twoside
\else
\fi
\advance\textheight by \topskip
\leftmargin\leftmargini
\labelwidth\leftmargini\advance\labelwidth-\labelsep

\begin{document}

\begin{center}
{\Large {\bf On Equivalence of Duffin-Kemmer-Petiau and Klein-Gordon Equations}}
\end{center}

\vspace{.5cm}

\begin{center}
{\large {V. Ya. Fainberg}}\footnote{%
Permanent address: P.N. Lebedev Institute of Physics, Moscow, Russia.}%
{\large {\ and B. M. Pimentel}}

\vspace{.3cm}

Instituto de F\'{\i}sica Te\'{o}rica\\[0pt]
Universidade Estadual Paulista\\[0pt]
Rua Pamplona 145\\[0pt]
01405-900 - S\~{a}o Paulo, S.P.\\[0pt]
Brazil\\[0pt]

\vspace{.5cm}

\begin{minipage}{340pt}
\centerline{\bf Abstract}
{A strict proof of equivalence between Duffin-Kemmer-Petiau (DKP) and Klein-Gordon 
(KG) theories is presented for physical S-matrix elements in the case of charged scalar 
particles interacting in minimal way with an external or  quantized electromagnetic field. 
First, Hamiltonian canonical approach to DKP theory is developed in both component 
and matrix form. The theory is then quantized through the construction of the 
generating functional for Green functions (GF) and the physical matrix elements of 
S-matrix are proved to be relativistic invariants. The equivalence between both 
theories is then proved using the connection between GF and the elements of 
S-matrix, including the case of only many photons states, and for more general 
conditions - so called reduction formulas of Lehmann, Symanzik, Zimmermann.}
\end{minipage}

\setcounter{footnote}{0}
\end{center}

\section{Introduction}

More than 60 years ago G. Petiau \cite{Petiau}, R Duffin \cite{Duffin} and
N. Kemmer \cite{Kemmer} proposed the first order equation (DKP equation) for
descripition of spin 0 and 1 particles. This period of time is
conventionally divided in three periods: the first one from 1939 until,
aproximately 1970; the second from 1970 to 1980 and the last one from1980 on.

During the first period the majority of the papers about DKP equation was
devoted to the development of DKP formalism and to the investigation of DKP
charged particles interaction with eletromagnetic field (EM field). For many
classes of processes (such as QE of spin 0 mesons, meso-atom and others)
calculations based on DKP and KG equations yield identical results \cite
{Akhiezer}, including one-loop corrections \cite{Umezawa,Akhiezer,Kinoshita}%
\footnote{%
A rich list of references with historical comments can be found in reference 
\cite{Nieto}. Unfortunately in this work there are no references to the
works by I. Gelfand and A. Yaglom, who obtained the first order equation for
particles with fixed arbitrary spin. For references to these and others
works see \cite{Ginzburg}.}.

A very important contribution to the understanding of this question was made
by A. Wightman in his paper \cite{Wightman} devoted to the 70th birthday of
P. A. M. Dirac in 1971. He showed that, for minimal coupling with EM field ($%
\sim e\overline{\psi }\beta _{\mu }\psi A^{\mu }$) in DKP theory, one has
stability of DKP equation for spin-0 particle under smooth local pertubation
of external field or, expressing in another way, that the retarded character
of DKP equation solutions is conserved under such pertubation as well as in
Dirac equation without anomalous magnetic moment.

The instability of such sort has been discussed by G. Velo and D. Zwanziger
in \cite{Velo}, where they showed that it destroys causality in
Rarita-Swinger equation for spin 3/2 particles (in external field).

The central point of the proof by A. Wightman is: the retarded character of
solution is connected with the renormalizability of the theory in the case
of interaction with a quantized EM field.

The second period can be characterized as by some disappointments and
hesitations. By this time two great discoveries had been made:\ parity
violation and creation of unified theory of eletro-weak interaction
(Weinberg-Salam theory or Standard Model). The question about the
equivalence of both DKP and KG theories arises again at the attempts to
describe new processes. Many works (see references in \cite{Nieto}) have
been made applying DKP formalism to decays of $K$ and other unstable mesons
and to strong interaction. The conclusion presented in reference \cite{Nieto}
was not optimistic:\ DKP formalism in some cases yield {\bf different}
results from a second order formalism\footnote{%
Moreover, in work \cite{Niet2} it is affirmed that DKP theory gives for $K$
meson decay qualitatively different results when compared to KG-formalism}.

The third period goes under the sign of uncertainness: are both DKP and KG
equivalent or not? Not so many papers have been published on this theme. In
our opinion one of main reasons for the decrease of interest in DKP
formalism in the last period is the conclusion about nonequivalence between
DKP and KG theories\footnote{%
See section 5; Conclusion, point 5.}.

We believe that the equivalence between these theories in the case of
nonstable particles can be proved as well as for all processes which are
described by renormalizable theories. This question, however, goes beyond
the scope of this paper.

As we know there are no strict proof of the equivalence between DKP and KG
theories in Quantum Electrodynamics of spin 0 particles, too. Coincident
results have been obtained for many processes in first order pertubation
theory, one loop corrections and the infrared approximation \cite{Akhiezer,
Kinoshita, pimentel}.

The main goal of this paper is to give a strict proof of the equivalence of
DKP and KG theories for charged scalar particles interacting with an
external or quantized EM field in minimal way for physical matrix elements
in any order of pertubation theory.

In section 2 Hamiltonian canonical approach to DKP theory is developed in
component and matrix forms. The construction of generating functional of
Green function (GF) of DKP theory is used for quantization of the theory and
the physical matrix elements of S-matrix are proved to be relativistic
invariants.

In section 3 the equivalence of DKP and KG theories is proved for physical
matrix elements utilizing the connection between GF and the elements of
S-matrix, including the case of only many photons states. In section 4 the
equivalence of both theories is shown in the more general framework
condition - so called reduction formulas of Lehmann, Symanzik, Zimmermann 
\cite{LSZ}.

In section 5 we shortly discuss the basic results and questions about
construction of renormalizable DKP theory for spin 0 particles.

The appendix contains some proofs.

\section{Canonical Quantization}

\subsection{Hamiltonian approach in component form}

Our aim is to construct the Hamiltonian for DKP theory which is one with
constraints due to degeneration of $\beta $\ matrices first we will work in
component representation. The Lagrangian density is 
\begin{equation}
{\cal L}=\overline{\psi }\left( i\beta _{\mu }D^{\mu }-m\right) \psi
\label{2.1}
\end{equation}
where $D^{\mu }=\partial ^{\mu }-ieA^{\mu }$; $\partial ^{\mu }=\frac{%
\partial }{\partial x_{\mu }}$; $g_{\mu \nu }=diag\left\{ 1,-1,-1,-1\right\} 
$. Primarily one considers $A^{\mu }$ as an external EM field. We choose the 
$\beta _{\mu }$ matrices in the following form: 
\begin{eqnarray}
\beta _{0} &=&\left| 
\begin{array}{ccccc}
0 & i & 0 & 0 & 0 \\ 
-i & 0 & 0 & 0 & 0 \\ 
0 & 0 & 0 & 0 & 0 \\ 
0 & 0 & 0 & 0 & 0 \\ 
0 & 0 & 0 & 0 & 0
\end{array}
\right| ;\ \ \beta _{1}=\left| 
\begin{array}{ccccc}
0 & 0 & -i & 0 & 0 \\ 
0 & 0 & 0 & 0 & 0 \\ 
-i & 0 & 0 & 0 & 0 \\ 
0 & 0 & 0 & 0 & 0 \\ 
0 & 0 & 0 & 0 & 0
\end{array}
\right|  \nonumber \\
\beta _{2} &=&\left| 
\begin{array}{ccccc}
0 & 0 & 0 & -i & 0 \\ 
0 & 0 & 0 & 0 & 0 \\ 
0 & 0 & 0 & 0 & 0 \\ 
-i & 0 & 0 & 0 & 0 \\ 
0 & 0 & 0 & 0 & 0
\end{array}
\right| ;\ \ \beta _{3}=\left| 
\begin{array}{ccccc}
0 & 0 & 0 & 0 & -i \\ 
0 & 0 & 0 & 0 & 0 \\ 
0 & 0 & 0 & 0 & 0 \\ 
0 & 0 & 0 & 0 & 0 \\ 
-i & 0 & 0 & 0 & 0
\end{array}
\right|  \label{2.2}
\end{eqnarray}
\begin{eqnarray}
\overline{\psi }_{\alpha } &=&\left( \psi ^{\ast }\eta \right) _{\alpha
}=\left( \varphi ^{\ast },\varphi ^{\ast 0},-\varphi ^{\ast 1},-\varphi
^{\ast 2},-\varphi ^{\ast 3}\right) ;\ \eta =2\left( \beta _{0}\right) ^{2}-1
\nonumber \\
\psi _{\alpha } &=&\left( \varphi ,\varphi ^{0},\varphi ^{1},\varphi
^{2},\varphi ^{3}\right) .  \label{2.3}
\end{eqnarray}

In component form ${\cal L}$ is then equal to 
\begin{equation}
{\cal L}=-\varphi ^{\ast }D_{\mu }\varphi ^{\mu }+\varphi ^{\ast \mu }D_{\mu
}\varphi -m\left( \varphi ^{\ast }\varphi +\varphi ^{\ast \mu }\varphi _{\mu
}\right)  \label{2.4}
\end{equation}

From equation (\ref{2.4}) one finds all the 1$^{st}$ stage constraints: 
\begin{equation}
p=\frac{\partial {\cal L}}{\partial \stackrel{.}{\varphi }}=\varphi ^{\ast
0},\ p^{\ast }=\frac{\partial {\cal L}}{\partial \stackrel{.}{\varphi }%
^{\ast }}=0  \label{2.5}
\end{equation}
\begin{equation}
p_{0}=\frac{\partial {\cal L}}{\partial \stackrel{.}{\varphi }^{0}}=-\varphi
^{\ast },\ p_{0}^{\ast }=\frac{\partial {\cal L}}{\partial \stackrel{.}{%
\varphi }^{\ast 0}}=0  \label{2.6}
\end{equation}
\begin{equation}
p_{i}=\frac{\partial {\cal L}}{\partial \stackrel{.}{\varphi }^{i}}=0,\
p_{i}^{\ast }=\frac{\partial {\cal L}}{\partial \stackrel{.}{\varphi }^{\ast
i}}=0,\ i=1,2,3.  \label{2.7}
\end{equation}

Now we construct the initial Hamiltonian $H$: 
\begin{eqnarray}
H=\int d^{3}x\left\{ p_{a}\stackrel{.}{\varphi }^{a}-{\cal L}\right\}
&=&\int d^{3}x\left\{ \varphi ^{\ast }\partial _{i}\varphi ^{i}-\varphi
^{\ast i}\partial _{i}\varphi +m\left( \varphi ^{\ast }\varphi +\varphi
^{\ast \mu }\varphi _{\mu }\right) \right.  \nonumber \\
&&\left. +ie\left( \varphi ^{\ast \mu }A_{\mu }\varphi -\varphi ^{\ast
}A_{\mu }\varphi ^{\mu }\right) \right\}  \label{2.8}
\end{eqnarray}

Here $p_{a}=\left\{ p,p^{\ast },p_{0},p_{0}^{\ast },p_{i},p_{i}^{\ast
}\right\} $ and we used equations (\ref{2.5}) to (\ref{2.7}).

From equations (\ref{2.5}) to (\ref{2.7}) we get the 1$^{st}$ stage
constraints\footnote{%
We follow terminology of the book \cite{Gitman}.}: 
\begin{equation}
\theta =p-\varphi ^{\ast 0},\ \theta ^{\ast }=p^{\ast },\ \theta
_{0}=p_{0}+\varphi ^{\ast },\ \theta _{0}^{\ast }=p_{0}^{\ast }  \label{2.9}
\end{equation}
\begin{equation}
\theta _{i}=p_{i},\ \theta _{i}^{\ast }=p_{i}^{\ast }.  \label{2.10}
\end{equation}

Thus one can speak that the initial $H$ is the Hamiltonian on the constraint
surface: 
\begin{equation}
H=\int d^{3}x\left\{ p_{a}\stackrel{.}{\varphi }^{a}-{\cal L}\right\}
_{\theta _{a}=0}.  \label{2.8a}
\end{equation}

It is easy to check that the Poisson brackets for constraints $\theta _{a}$
are\footnote{%
All variables: $p_{a},\varphi _{a}$\ depend on {\bf x.}Therefore the r.h.s.
of equation (12) is $\sim \delta ({\bf x}-{\bf y})$ which we omited.}: 
\begin{equation}
\left\{ \theta ,\theta _{0}^{\ast }\right\} =-1;\ \left\{ \theta ^{\ast
},\theta _{0}\right\} =-1.  \label{2.11}
\end{equation}

All others brackets equal zero. Constraints (\ref{2.9}) are second-class
constraints\footnote{%
For simplicity we will not use the Dirac brackets.} as they do not disappear
on the constraint surface $\theta _{\alpha }=0$.

From equation (\ref{2.11}) it follows that 
\begin{equation}
rank\left\| \theta _{a},\theta _{b}\right\| =4.  \label{2.12}
\end{equation}

This means that if we introduce new hamiltonian off constraint surface for
definition Lagrangian multipliers $\lambda _{a}$ 
\begin{equation}
H_{2}^{\left( 1\right) }=H+\int d^{3}x\lambda _{a}\left( {\bf x}\right)
\theta ^{a}\left( {\bf x}\right) ,  \label{2.13}
\end{equation}
and demanding that all constraints conserve in time 
\begin{equation}
\stackrel{.}{\theta }^{a}=\left\{ \theta ^{a},H_{2}^{\left( 1\right)
}\right\} =0,  \label{2.14}
\end{equation}
we can not define all $\lambda _{a}$, but only $\lambda $, $\lambda ^{\ast }$%
, $\lambda _{0}$, $\lambda _{0}^{\ast }$.

Six constraints from (\ref{2.14}) give the six 2$^{nd}$ stage ones that are: 
\begin{eqnarray}
\stackrel{.}{\theta }_{i} &=&\left\{ \theta _{i},H_{2}^{\left( 1\right)
}\right\} =\left\{ \theta _{i},H\right\} =-\frac{\partial H}{\partial
\varphi ^{i}}=D_{i}^{\ast }\varphi ^{\ast }-m\varphi _{i}^{\ast }\equiv
\theta _{i}^{2},  \label{2.15a} \\
\stackrel{.}{\theta }_{i}^{\ast } &=&\left\{ \theta _{i}^{\ast
},H_{2}^{\left( 1\right) }\right\} =-\frac{\partial H}{\partial \varphi
^{\ast i}}=D_{i}\varphi -m\varphi _{i}\equiv \theta _{i}^{\ast 2}.
\label{2.15}
\end{eqnarray}

Including all constraints (ten 1$^{st}$ stage and six 2$^{nd}$ stage) in
total system, where now 
\begin{equation}
\theta _{a}=\left\{ \theta ,\theta ^{\ast },\theta _{0},\theta _{0}^{\ast
},\theta _{i},\theta _{i}^{\ast },\theta _{i}^{2},\theta _{i}^{\ast
2}\right\} ,  \label{2.16}
\end{equation}
we can check that 
\begin{equation}
\func{rank}\left\| \theta _{a},\theta _{b}\right\| =16  \label{2.17}
\end{equation}
and 
\begin{equation}
\det \left\| \theta _{a},\theta _{b}\right\| =1.  \label{2.18}
\end{equation}

All these constraints are second class ones. According to the general scheme 
\cite{Gitman}\ we can find all $\lambda _{a}$ from the conditions: 
\begin{equation}
\stackrel{.}{\theta }_{a}=\left\{ \theta _{a},H_{2}^{\left( 1\right)
}\right\} =0.  \label{2.20}
\end{equation}

From equations (\ref{2.9}), (\ref{2.10}), (\ref{2.13}), (\ref{2.15}), (\ref
{2.16}), and (\ref{2.20}) one gets: 
\begin{equation}
\lambda =m\varphi _{0}+ieA_{0}\varphi ;\ \lambda ^{\ast }=m\varphi
_{0}^{\ast }-ieA_{0}\varphi ^{\ast };  \label{2.21}
\end{equation}
\begin{equation}
\lambda _{0}=-\frac{1}{m}\left( D_{i}D^{i}\varphi +m^{2}\varphi \right)
+ie\varphi _{0}A_{0};\ \lambda _{0}^{\ast }=-\frac{1}{m}\left( D_{i}^{\ast
}D^{\ast i}\varphi ^{\ast }+m^{2}\varphi ^{\ast }\right) -ie\varphi
_{0}^{\ast }A_{0};  \label{2.22}
\end{equation}
\begin{equation}
\lambda _{i}=-\frac{1}{m}D_{i}\left( m\varphi _{0}+ieA_{0}\varphi \right) ;\
\lambda _{i}^{\ast }=-\frac{1}{m}D_{i}^{\ast }\left( m\varphi _{0}^{\ast
}+ieA_{0}\varphi ^{\ast }\right) .  \label{2.23}
\end{equation}

Equation (\ref{2.23}) is a consequence of conservation of the 2$^{nd}$ stage
constraints: 
\begin{equation}
\theta _{i}^{2}\equiv \stackrel{.}{\theta }_{i}=\left\{ \theta
_{i},H_{2}^{\left( 1\right) }\right\} =0;\ \theta _{i}^{\ast 2}\equiv 
\stackrel{.}{\theta }_{i}^{\ast }=\left\{ \theta _{i},H_{2}^{\left( 1\right)
}\right\} .  \label{2.24}
\end{equation}

The main criterium of the correctness of the canonical or $H$-approach to
the ${\cal L}$-theory is the coincidence of the Lagrangian and Hamiltonian
equations of motion.

To get $H$-equations of motion we must introduce solutions (\ref{2.21}) to (%
\ref{2.23}) in equation (\ref{2.13}) and to consider the following equations:

\begin{eqnarray}
\stackrel{.}{\varphi } &=&\left\{ \varphi \left( {\bf x}\right)
,H_{2}^{\left( 1\right) }\right\} =\int d^{3}y\left\{ \varphi \left( {\bf x}%
\right) ,\lambda ^{a}\left( {\bf y}\right) \theta _{a}\left( {\bf y}\right)
\right\}  \nonumber \\
&=&\int d^{3}y\lambda \left( {\bf y}\right) \left\{ \varphi \left( {\bf x}%
\right) ,\theta \left( {\bf y}\right) \right\} =\lambda \left( {\bf x}%
\right) =m\varphi _{0}+ieA_{0}\varphi  \label{2.25}
\end{eqnarray}
or 
\begin{equation}
D_{0}\varphi -m\varphi _{0}=0,  \label{2.26a}
\end{equation}
and analogously 
\begin{equation}
D_{0}^{\ast }\varphi ^{\ast }-m\varphi _{0}^{\ast }=0.  \label{2.26b}
\end{equation}

\ Conservation of constraint $\theta _{i}$; $\stackrel{.}{\theta }%
_{i}=\left\{ \theta _{i},H_{2}^{\left( 1\right) }\right\} =0$; gives: 
\begin{equation}
D^{i}\varphi -m\varphi ^{i}=0;\ D^{\ast i}\varphi ^{\ast }-m\varphi ^{\ast
i}=0.  \label{2.27}
\end{equation}

Collecting equations (\ref{2.26a}), (\ref{2.26b}) and (\ref{2.27}) one gives 
\begin{equation}
D^{\mu }\varphi -m\varphi ^{\mu }=0;\ D^{\ast \mu }\varphi ^{\ast }-m\varphi
^{\ast \mu }=0.  \label{2.28}
\end{equation}

Consider 
\begin{eqnarray*}
\stackrel{.}{\varphi }_{0}\left( {\bf x}\right) &=&\left\{ \varphi
_{0},H_{2}^{\left( 1\right) }\right\} =\int d^{3}y\lambda ^{0}\left( {\bf y}%
\right) \left\{ \varphi _{0}\left( {\bf x}\right) ,\theta _{0}\left( {\bf y}%
\right) \right\} = \\
&=&\lambda ^{0}\left( {\bf x}\right) =-\frac{1}{m}\left( D_{i}D^{i}\varphi
+m^{2}\varphi \right) +ie\varphi _{0}A^{0}= \\
&=&-\frac{1}{m}D^{i}\left( D_{i}\varphi +m\varphi _{i}\right) -\left(
D_{i}\varphi ^{i}+m\varphi \right) +ie\varphi _{0}A^{0}= \\
&=&-\left( D_{i}\varphi ^{i}+m\varphi \right) +ie\varphi _{0}A^{0}
\end{eqnarray*}
being the first term equal to zero due to equation (\ref{2.27}). Or 
\begin{equation}
D_{\mu }\varphi ^{\mu }+m\varphi =0.  \label{2.29}
\end{equation}

We obtain all ${\cal L}$-equations, which follow from equation (\ref{2.4}).

Thus we get by canonical way, from equations (\ref{2.28}) and (\ref{2.29}), 
{\bf on the classical level}, the KG equation for $\varphi $-field: 
\begin{equation}
D_{\mu }D^{\mu }\varphi +m\varphi =0.  \label{2.30}
\end{equation}

Obviously, this result follows in a very simple way from component form of $%
{\cal L}$, equations (\ref{2.4}), too.

\subsection{Quantization. Generating Functional}

For canonical quantization of the theory with second class constraints%
\footnote{%
Quantization of theories with the 2$^{\text{nd \ }}-$class constraints
devouted many papers, beginning with classical work of Dirac \cite{Dirac},
see also \cite{Fradkin, Faddeev}.} \cite{Gitman, Dirac} one chooses the new
canonical variables:

\begin{eqnarray}
\omega ^{\left( i\right) } &=&\varphi ,\omega ^{\left( 2\right) }=p;\Omega
^{\left( 1\right) }=-\theta =\varphi ^{0}-p,\Omega ^{\left( 2\right)
}=\theta _{0}=p_{0};\Omega _{i}^{\left( 1\right) }=-\theta _{i}^{\left(
2\right) }=\varphi _{i}-\frac{D_{i}\varphi }{m},  \nonumber \\
\omega _{0}^{\left( i\right) } &=&\varphi ,\omega _{0}^{\left( 2\right)
}=p;\Omega ^{\left( 1\right) }=\theta _{0}=\varphi ^{\ast }+p_{0},\Omega
_{0}^{\left( 2\right) }=\theta ^{\ast }=p^{\ast };\Omega _{i}^{\left(
2\right) }=-\theta _{i}^{\ast \left( 2\right) }=\varphi _{i}^{\ast }-\frac{%
D_{i}^{\ast }\varphi ^{\ast }}{m},  \nonumber \\
\Omega _{i}^{\left( 1\right) } &=&\theta _{i}=p_{i}  \nonumber \\
\Omega _{i}^{\left( 1\right) } &=&\theta _{i}^{\ast }=p_{i}^{\ast }
\label{2.31}
\end{eqnarray}

Here we have four physical variables, $\omega ^{\left( k\right) }$ and $%
\omega _{0}^{\left( k\right) }$ , $k=1,2,$ and sixteen constraints $\Omega
_{a}^{\left( k\right) }=0$, which allow to express the rest variables
through $\ \omega ^{\left( k\right) },\omega _{0}^{\left( k\right) }$.

Now the rules of quantization are very simple \cite{Gitman, Dirac}: 
\begin{equation}
\left. 
\begin{array}{c}
\left[ \stackrel{\symbol{94}}{\varphi }\left( {\bf x}\right) ,\stackrel{%
\symbol{94}}{p}\left( {\bf y}\right) \right] =i\left\{ \varphi \left( {\bf x}%
\right) ,p\left( {\bf y}\right) \right\} =i\delta \left( {\bf x}-{\bf y}%
\right) \\ 
\left[ \stackrel{\symbol{94}}{\varphi }_{0}\left( {\bf x}\right) ,\stackrel{%
\symbol{94}}{p}_{0}\left( {\bf y}\right) \right] =i\left\{ \varphi
_{0}\left( {\bf x}\right) ,p_{0}\left( {\bf y}\right) \right\} =i\delta
\left( {\bf x}-{\bf y}\right) \\ 
\stackrel{\symbol{94}}{\Omega }_{a}^{\left( k\right) }=0.
\end{array}
\right\}  \label{2.31a}
\end{equation}
In terms of $\stackrel{\symbol{94}}{p},$ $\stackrel{\symbol{94}}{\varphi }$, 
$\stackrel{\symbol{94}}{p}_{0}$, $\stackrel{\symbol{94}}{\varphi }_{0}$ the
Dirac brackets coincide with the Poisson ones.

Then we get for Heisenberg operators $\stackrel{\symbol{94}}{\varphi }$ the
equation (\ref{2.30}) and for $H$ the same expression as in KG theory.
However, for proof of equvalence of the physical matrix elements of S-matrix
for scalar particles and of many photons GF in both theories it is more
simple to start not from $\stackrel{\symbol{94}}{H}$ , but from generating
functional for GF of DKP theory, wich follows from canonical
quantization,too.

For generating functional of DKP theory in component form we get

\begin{eqnarray}
{\cal Z}\left( {\cal J},{\cal J}^{\ast },{\cal J}_{\mu },{\cal J}_{\mu
}^{\ast }\right) &=&{\cal Z}_{0}^{-1}\int {\cal D}\varphi _{a}{\cal D}%
p_{a}\mu \left( \theta ^{a}\right)  \nonumber \\
&&\times \exp \left\{ i\int d^{4}x\left( p_{a}\stackrel{.}{\varphi }_{a}-%
{\cal H}_{2}^{\left( 1\right) }+\varphi _{a}{\cal J}^{a}\right) \right\} ,
\label{2.32}
\end{eqnarray}
where 
\begin{equation}
\mu \left( \theta ^{a}\right) =\stackunder{a}{\sqcap }\delta \left( \theta
_{a}\right) \left( \det \left\| \theta _{a},\theta _{b}\right\| \right)
^{1/2},\qquad {\cal Z}_{0}={\cal Z}\left( 0,0,0,0\right) ,\ {\cal H}%
_{2}^{\left( 1\right) }=\text{density of }H_{2}^{\left( 1\right) }.
\label{2.33}
\end{equation}

Taking into account equations (\ref{2.9}), (\ref{2.10}), (\ref{2.13}), (\ref
{2.15a}), (\ref{2.15}), (\ref{2.33}) and integrating over all\ momentum $%
p_{a}$ we get 
\begin{eqnarray}
{\cal Z}\left( {\cal J},{\cal J}^{\ast },{\cal J}_{\mu },{\cal J}_{\mu
}^{\ast }\right) &=&{\cal Z}_{0}^{-1}\stackunder{a}{\sqcap }\int {\cal D}%
\varphi _{a}\stackunder{i=1}{\stackrel{3}{\sqcap }}\delta \left(
D^{i}\varphi -m\varphi ^{i}\right) \delta \left( D^{\ast i}\varphi ^{\ast
}-m\varphi ^{\ast i}\right)  \nonumber \\
&&\times \exp \left\{ i\int d^{4}x\left( \varphi ^{\ast \mu }D_{\mu }\varphi
-\varphi ^{\ast }D_{\mu }\varphi ^{\mu }-m\left( \varphi ^{\ast }\varphi
+\varphi ^{\ast \mu }\varphi _{\mu }\right) \right. \right.  \nonumber \\
&&\left. \left. +{\cal J}^{\ast }\varphi +{\cal J}\varphi ^{\ast }+{\cal J}%
_{\mu }\varphi ^{\ast \mu }+{\cal J}_{\mu }^{\ast }\varphi ^{\mu }\right)
\right\} .  \label{2.34}
\end{eqnarray}

The difference between $H$-quantization of DKP-theory and formal ${\cal L}$%
-quan\-ti\-za\-tion consists in appearance in equation (\ref{2.34}) of two
functional $\delta $-functions, which reflect the existence of the 2$^{nd}$
stage constraints (\ref{2.15a}) and (\ref{2.15}) in $H$- approach to DKP
theory.

After integration in equation (\ref{2.34}) over $\varphi _{a}$ and utilizing
the $\delta $-function we get: 
\begin{eqnarray}
{\cal Z}\left( {\cal J},{\cal J}^{\ast },{\cal J}_{\mu },{\cal J}_{\mu
}^{\ast }\right) &=&\exp \left\{ i\int d^{4}xd^{4}y\left( m{\cal J}^{\ast
}\left( x\right) G\left( x,y\right) {\cal J}\left( y\right) \right. \right. 
\nonumber \\
&&\left. -{\cal J}^{\ast }\left( x\right) G\left( x,y\right) D_{\mu }{\cal J}%
^{\mu }\left( y\right) -{\cal J}\left( x\right) \left( D_{\mu }^{\ast }{\cal %
J}^{\ast \mu }\left( y\right) \right) G\left( x,y\right) \right.  \nonumber
\\
&&\left. -\frac{1}{m}{\cal J}^{\mu }\left( x\right) D_{\mu }G\left(
x,y\right) D_{\nu }{\cal J}^{\nu }\left( y\right) \right.  \nonumber \\
&&\left. -\frac{1}{m}{\cal J}_{0}\left( x\right) \delta ^{4}\left(
x-y\right) {\cal J}_{0}\left( y\right) \right) .  \label{2.35}
\end{eqnarray}

Here 
\begin{equation}
G\left( x,y\right) =\left( D_{\mu }D^{\mu }+m^{2}\right) ^{-1}\delta
^{4}\left( x-y\right)  \label{2.35a}
\end{equation}
is the GF of the scalar charged field $\varphi $ in external EM field.

If one formally makes ${\cal J}_{\mu }={\cal J}_{\mu }^{\ast }=0$ we obtain
from equation (\ref{2.35}) the generating functional GF in KG theory.

In the general case, as it will be shown bellow (see equation (\ref{2.47})),
the generating functional (\ref{2.35}) exactly coincides with that
calculated in matrix form, where 
\begin{equation}
\stackrel{\_}{I}_{\alpha }=\left( {\cal J}^{\ast },{\cal J}_{0}^{\ast 0},-%
{\cal J}^{\ast 1},-{\cal J}^{\ast 2},-{\cal J}^{\ast 3}\right) ;\ I_{\alpha
}=\left( {\cal J},{\cal J}_{0},{\cal J}^{1},{\cal J}^{2},{\cal J}^{3}\right)
.  \label{2.36}
\end{equation}

We want stress that det $\left( D_{\mu }D^{\mu }+m^{2}\right) ^{-1}$does not
appear in equation (\ref{2.35}), see point 2 after equation (\ref{2.48}).

\subsection{H-approach. Matrix form}

Starting from equation (\ref{2.1}) we can define the momenta 
\begin{equation}
p_{\alpha }=\frac{\partial {\cal L}}{\partial \stackrel{.}{\psi }_{\alpha }}%
=i\left( \psi ^{\ast }\beta _{0}\right) _{\alpha },\ p_{\alpha }^{\ast }=%
\frac{\partial {\cal L}}{\partial \stackrel{.}{\psi }_{\alpha }^{\ast }}=0,
\label{2.37}
\end{equation}
\begin{equation}
p^{\alpha }=\frac{\partial {\cal L}}{\partial \stackrel{.}{\psi }_{\alpha
}^{\alpha }}=0,\ p^{\ast \alpha }=\frac{\partial {\cal L}}{\partial 
\stackrel{.}{\psi }_{\alpha }^{\ast }}=0,\ \text{for }\alpha =1,2,3\equiv i.
\label{2.38}
\end{equation}

The initial Hamiltonian $H$ is equal to: 
\begin{eqnarray}
H &=&\int d^{3}x\left\{ p^{\alpha }\stackrel{.}{\psi }_{\alpha }+\ p^{\ast
\alpha }\stackrel{.}{\psi }_{\alpha }^{\ast }-{\cal L}\right\} =  \nonumber
\\
&=&\int d^{3}x\left( -i\overline{\psi }\beta _{k}D^{k}\psi +\overline{\psi }%
m\psi -e\overline{\psi }\beta _{0}A^{0}\psi \right)  \nonumber \\
&=&\int d^{3}x\left( -i\overline{\psi }\beta _{k}\partial ^{k}\psi +m%
\overline{\psi }\psi -e\overline{\psi }\beta _{\mu }A^{\mu }\psi \right) ,\
k=1,2,3  \label{2.39}
\end{eqnarray}

This is exactly equation (\ref{2.8}), if we write equation (\ref{2.39}) in
component form.

Here we write down all the 1$^{st}$ stage and the 2$^{st}$ stage constraints
and Lagrangian multipliers in matrix form, omitting calculations:

- 1$^{st}$ stage constraints 
\begin{eqnarray}
\theta _{\alpha } &=&p_{\alpha }-i\left( \overline{\psi }\beta _{0}\right)
_{\alpha }  \label{2.40} \\
\theta _{\alpha }^{\ast } &=&p_{\alpha }^{\ast }  \label{2.41}
\end{eqnarray}

- 2$^{nd}$ stage constraints 
\begin{equation}
\left. 
\begin{array}{c}
\theta _{2}^{\alpha }=\left[ \left( 1-\left( \beta _{0}\right) ^{2}\right)
\left( i\beta _{k}D^{k}-m\right) \psi \right] ^{\alpha } \\ 
\theta _{2}^{\ast \alpha }=\left[ \overline{\psi }\left( i\beta _{k}%
\overleftarrow{D^{\ast }}^{k}+m\right) \left( 1-\left( \beta _{0}\right)
^{2}\right) \right] ^{\alpha }
\end{array}
\right\}  \label{2.42}
\end{equation}

- Lagrangian multipliers 
\begin{equation}
\left( \beta _{0}^{2}\lambda \right) _{\alpha }=\left[ i\beta _{0}\left(
i\beta _{k}D^{k}-m\right) \psi \right] _{\alpha };\ \left( \lambda ^{\ast
}\beta _{0}^{2}\right) _{\alpha }=-\left[ \overline{\psi }\left( i\beta _{k}%
\overleftarrow{D}^{\ast k}+m\right) \beta _{0}\right] _{\alpha };
\label{2.43}
\end{equation}
\begin{eqnarray}
\left[ \left( 1-\left( \beta _{0}\right) ^{2}\right) \left( m-i\beta
_{k}D^{k}\right) \eta \right] _{\alpha \beta }\lambda ^{\beta } &=&0, 
\nonumber \\
\lambda ^{\ast \beta }\left[ \eta \left( i\beta _{k}\overleftarrow{D}^{\ast
k}+m\right) \left( 1-\left( \beta _{0}\right) ^{2}\right) \right] _{\beta
\alpha } &=&0.  \label{2.44}
\end{eqnarray}

All these $\lambda _{\alpha }$,$\lambda _{\alpha }^{\ast }$ expressed
through the components coincide with the corresponding $\lambda _{a}$ in
equations (\ref{2.21}) to (\ref{2.23}).

\subsection{Quantization. Generating Functional.Matrix Form}

As well as in component form in equation (\ref{2.32}) we get, for generating
functional in external EM field, the following expression: 
\begin{eqnarray}
{\cal Z}\left( I,\stackrel{\_}{I}\right) &=&{\cal Z}_{0}^{-1}\int {\cal D}%
\psi {\cal D}\overline{\psi }\delta \left( \left( 1-\left( \beta _{0}\right)
^{2}\right) \left( i\beta _{k}D^{k}-m\right) \psi \right)  \nonumber \\
&&\times \delta \left( \overline{\psi }\left( i\beta _{k}\overleftarrow{%
D^{\ast }}^{k}+m\right) \left( 1-\left( \beta _{0}\right) ^{2}\right) \right)
\nonumber \\
&&\times \exp \left\{ i\int d^{4}x\left( \overline{\psi }\left( i\beta _{\mu
}D^{\mu }-m\right) \psi +\stackrel{\_}{I}\psi +\overline{\psi }I\right)
\right\}  \label{2.45}
\end{eqnarray}

Introducing the auxiliary fields $C$ and $\stackrel{\_}{C}$ instead
functional $\delta -$ function in equation (\ref{2.45}) we get 
\begin{eqnarray}
{\cal Z}\left( I,\stackrel{\_}{I}\right) &=&{\cal Z}_{0}^{-1}\int {\cal D}%
\psi {\cal D}\overline{\psi }{\cal D}C{\cal D}\stackrel{\_}{C}\times \exp
\left\{ i\int d^{4}x\left( \overline{\psi }\left( i\beta _{\mu }D^{\mu
}-m\right) \psi \right. \right.  \nonumber \\
&&\left. +\stackrel{\_}{C}\left( 1-\left( \beta _{0}\right) ^{2}\right)
\left( i\beta _{\mu }D^{\mu }-m\right) \psi \right.  \nonumber \\
&&\left. \left. \overline{\psi }\left( i\beta _{\mu }\overleftarrow{D^{\ast }%
}^{\mu }+m\right) \left( 1-\left( \beta _{0}\right) ^{2}\right) C+\stackrel{%
\_}{I}\psi +\overline{\psi }I\right) \right\} ,  \label{2.46}
\end{eqnarray}
where one used that $\beta _{0}\left( 1-\left( \beta _{0}\right) ^{2}\right)
=0$.

Integrating over all fields $\psi $, $\overline{\psi }$, $C$ and $\stackrel{%
\_}{C}$\ we finally obtain: 
\begin{eqnarray}
{\cal Z}\left( I,\stackrel{\_}{I}\right) &=&\exp \left\{ -i\int
d^{4}xd^{4}yI\left( x\right) \left( S\left( x,y,A\right) \right. \right. 
\nonumber \\
&&\left. \left. +\frac{1}{m}\left( 1-\left( \beta _{0}\right) ^{2}\right)
\delta ^{4}\left( x-y\right) \right) I\left( y\right) \right\} ,
\label{2.47}
\end{eqnarray}
where we have introduced the total GF of DK particle in external EM field $%
A_{\mu }$: 
\begin{equation}
S\left( x,y,A\right) =\left( i\beta _{\mu }D^{\mu }-m\right) ^{-1}\delta
^{4}\left( x-y\right)  \label{2.48}
\end{equation}

One can make some important comments about expression (\ref{2.48}).

1)\ If one writes down equation (\ref{2.48}) in component form we get
equation (\ref{2.35}).

2) When we integrate over $\psi $, $\overline{\psi }$, $C$ and $\stackrel{\_%
}{C}$ divergences appear and infinite expression for $\det \left( i\beta
_{\mu }D^{\mu }-m\right) ^{-1}$. All this multipliers also arise in ${\cal Z}%
_{0}$ and disappear from the final equation (\ref{2.47}) .

3) Nonrelativistic invariant term $\ \ \sim \left( 1-\left( \beta
_{0}\right) ^{2}\right) $ in equation (\ref{2.47}) arises at excluding
nonphysical component $\psi $, due to \ the second stage constraints ($%
\theta _{i}^{2}$, $\theta _{i}^{\ast 2}$ in equations (\ref{2.15a}) and (\ref
{2.15})).

We will show that this term (which does not depend on charge) does not
contribute to physical matrix elements of $S$-matrix.(see Appendix, point 1).

4)If one generalizes equations (\ref{2.45}) to (\ref{2.48}) to the case of
interaction of DK particles with quantized EM fields we get the following
expression for generating functional for all GF of the theory (in $\alpha $%
-gauge): 
\begin{eqnarray}
{\cal Z}\left( I,\stackrel{\_}{I},{\cal J}_{\mu }\right) &=&{\cal Z}%
_{0}^{-1}\int {\cal D}A_{\mu }\exp \left\{ -i\int d^{4}x\left( Tr\ln \frac{%
S\left( x,x,A\right) }{S\left( x,x,0\right) }-\frac{1}{4}F_{\mu \nu }F^{\mu
\nu }\right. \right.  \nonumber \\
&&\left. -{\cal J}_{\mu }A^{\mu }-\frac{1}{2\alpha }\left( \partial _{\mu
}A^{\mu }\right) ^{2}-\int d^{4}y\stackrel{\_}{I}\left( x\right) \left(
S\left( x,y,A\right) \right. \right.  \nonumber \\
&&\left. \left. \left. +\frac{1}{m}\left( 1-\left( \beta _{0}\right)
^{2}\right) \delta ^{4}\left( x-y\right) \right) I\left( y\right) \right)
\right\} ,  \label{2.49}
\end{eqnarray}

Here we insert in denominator and in ${\cal Z}_{0}$ infinit constant $\left(
\det S\left( x,x,0\right) \right) ^{-1}$. As it is well known, the term $\
\sim Tr\ln S\left( x,x,A\right) $ in equation (\ref{2.49}) is responsible
for appearance of all vacuum polarizations diagrams (see Section 3).

\section{Equivalence matrix elements of S-matrix for physical states in DKP
and KG Theories}

1)\ From the beginning we write down \ the physical operator's solution of
DK equation for {\bf free} particle; which will be needed for the proof of
equivalence of the both theories.

This solution can be written in the following form (see \cite{Umezawa}) 
\begin{equation}
\psi _{\alpha }^{(0)}\left( x\right) =\frac{1}{\left( 2\pi \right) ^{3/2}}%
\int d^{3}p\left\{ a^{-}\left( {\bf p}\right) e^{-ipx}\psi _{\alpha
}^{a}\left( p\right) +b^{+}\left( {\bf p}\right) e^{ipx}\psi _{\alpha
}^{b}\left( p\right) \right\} ,  \label{3.1}
\end{equation}
\begin{equation}
\overline{\psi }_{\alpha }^{(0)}\left( x\right) =\frac{1}{\left( 2\pi
\right) ^{3/2}}\int d^{3}p\left\{ a^{+}\left( {\bf p}\right) e^{ipx}%
\overline{\psi }_{\alpha }^{a}\left( p\right) +b^{-}\left( {\bf p}\right)
e^{-ipx}\overline{\psi }_{\alpha }^{b}\left( p\right) \right\} .  \label{3.2}
\end{equation}

Here

\begin{equation}
\left. 
\begin{array}{c}
\left( \stackrel{\wedge }{p}-m\right) \psi ^{a,b}=0;\ \overline{\psi }%
^{a,b}\left( \stackrel{\wedge }{p}-m\right) =0;\ \widehat{p}\equiv \beta
_{\mu }p^{\mu } \\ 
\left( i\beta _{\mu }\partial ^{\mu }-m\right) \psi ^{0}\left( x\right) =0;\
p_{0}=\omega \left( {\bf p}\right) =+\left( {\bf p}^{2}+m^{2}\right)
^{1/2}=\omega
\end{array}
\right\}  \label{3.3}
\end{equation}

Operators $a^{+}$, $a^{-}$, $b^{+}$and $b^{-}$ satisfy to the usual
commutation relations; 
\begin{equation}
\left. 
\begin{array}{c}
\psi _{\alpha }^{a}\left( p\right) =\sqrt{\frac{m}{2\omega }}\left( 1,-\frac{%
i\omega }{m},-\frac{ip^{1}}{m},-\frac{ip^{2}}{m},-\frac{ip^{3}}{m}\right) \\ 
\psi _{\alpha }^{b}\left( p\right) =\sqrt{\frac{m}{2\omega }}\left( 1,\frac{%
i\omega }{m},+\frac{ip^{1}}{m},+\frac{ip^{2}}{m},+\frac{ip^{3}}{m}\right)
\end{array}
\right\}  \label{3.4}
\end{equation}

It is easy to check the scalar products 
\begin{equation}
\left. 
\begin{array}{c}
\left( \psi ^{a}\left( p\right) ,\psi ^{a}\left( p\right) \right) \equiv
\psi _{\alpha }^{\ast a}\left( p\right) \left( \beta _{0}\right) _{\alpha
\beta }\psi _{\beta }^{a}\left( p\right) =\overline{\psi }^{a}\left(
p\right) \beta _{0}\psi ^{a}\left( p\right) =1 \\ 
\left( \psi ^{b}\left( p\right) ,\psi ^{b}\left( p\right) \right) \equiv
\psi _{\alpha }^{\ast b}\left( p\right) \left( \beta _{0}\right) _{\alpha
\beta }\psi _{\beta }^{b}\left( p\right) =\overline{\psi }^{b}\left(
p\right) \beta _{0}\psi ^{b}\left( p\right) =-1 \\ 
\left( \psi ^{a}\left( p\right) ,\psi ^{b}\left( p\right) \right) \equiv 
\overline{\psi }^{a}\left( p\right) \beta _{0}\psi ^{b}\left( p\right) =0
\end{array}
\right\}  \label{3.5}
\end{equation}

Thus, there are only two linearly independent (physical) solutions of free
DK equation.

One introduce 
\begin{equation}
\psi _{p}^{a,b}\left( x\right) =\pm \left( 2\pi \right) ^{-3/2}e^{\mp
ipx}\psi ^{a,b}\left( p\right) ,\ p_{0}=\omega .  \label{3.6}
\end{equation}

Then, utilizing equations (\ref{3.1}) to (\ref{3.5}) we get 
\begin{equation}
a^{-}\left( {\bf p}\right) =\int d^{3}x\psi _{p}^{\ast a}\left( x\right)
\beta _{0}\psi ^{0}\left( x\right) ;\ b^{+}\left( {\bf p}\right) =\int
d^{3}x\psi _{p}^{\ast b}\left( x\right) \beta _{0}\psi ^{0}\left( x\right) ,
\label{3.7}
\end{equation}
and $a^{+}=\left( a^{-}\right) ^{\ast }$, $b^{-}=\left( b^{+}\right) ^{\ast
} $.

2) Now we write the general connection between GF and matrix elements of $S$%
-matrix for physical states in DKP theory.

By definition $n$-particles GF of DKP-particles without \ external photon's
states equal, from equation (\ref{2.48}), to 
\begin{eqnarray}
G_{n}\left( x_{1},...,x_{n};y_{1},...,y_{n}\right) &=&\left( -1\right)
^{n}\left. \frac{\delta ^{2n}\ln {\cal Z}\left( I,\stackrel{\_}{I},{\cal J}%
_{\mu }=0\right) }{\delta \stackrel{\_}{I}\left( x_{1}\right) ...\delta 
\stackrel{\_}{I}\left( x_{n}\right) \delta I\left( y_{1}\right) ...\delta
I\left( y_{n}\right) }\right| _{I,\stackrel{\_}{I}=0}  \nonumber \\
&=&\left\langle 0\right| T\psi \left( x_{1}\right) ...\psi \left(
x_{n}\right) \overline{\psi }\left( y_{1}\right) ...\overline{\psi }\left(
y_{n}\right) \left| 0\right\rangle ,  \label{3.8}
\end{eqnarray}
where we omit nonconnected GF.

On the other hand, GF (\ref{3.8}) can be (formally) expressed through matrix
element of $S$-matrix in the following way: 
\begin{eqnarray}
G_{n}\left( x_{1},...,x_{n};y_{1},...,y_{n}\right) &=&\stackunder{i=1}{%
\stackrel{n}{\sqcap }}\int dx_{i}^{\prime }dy_{i}^{\prime }S\left(
x_{i}-x_{i}^{\prime }\right)  \nonumber \\
&&\times \Gamma _{n}\left( x_{1},...,x_{n};y_{1},...,y_{n}\right) S\left(
y_{i}^{\prime }-y_{i}\right)  \label{3.9}
\end{eqnarray}

Where $S\left( x-y\right) $ is the total one particle GF and the so-called $%
\Gamma _{n}$ functions which are equal, in momentum representation, to $%
S_{n}\left( p_{1},...,p_{n};q_{1},...,q_{n}\right) $, $n$-particles matrix
element of $S$-matrix for physical states on the mass-shell 
\begin{equation}
\left. 
\begin{array}{c}
\Gamma _{n}\left( p_{1},...,p_{n};q_{1},...,q_{n}\right) =S_{n}\left(
p_{1},...,p_{n};q_{1},...,q_{n}\right) \\ 
p_{i0}=\left( \left( {\bf p}_{i}\right) ^{2}+m^{2}\right) ^{1/2};\
q_{i0}=\left( \left( {\bf q}_{i}\right) ^{2}+m^{2}\right) ^{1/2};\
\sum\limits_{i=1}^{n}\left( q_{i}-p_{i}\right) =0
\end{array}
\right\}  \label{3.10}
\end{equation}

It is well known that one particle GF after renormalization has a pole $\sim
\left( m-\widehat{p}-i\varepsilon \right) ^{-1}$ with residue equal to one
and $\delta ^{4}\left( x-y\right) $ pecularity (see equation (\ref{2.47})).
Thus, if we consider the following physical matrix element 
\begin{equation}
A\equiv \lim \prod\limits\Sb i=1  \\ p_{i0}\rightarrow \left( \left( {\bf p}%
_{i}\right) ^{2}+m^{2}\right) ^{1/2}  \\ q_{i0}\rightarrow \left( \left( 
{\bf q}_{i}\right) ^{2}+m^{2}\right) ^{1/2}  \endSb ^{n}\overline{\psi }%
^{a,b}\left( p_{i}\right) \left( \overrightarrow{\stackrel{\wedge }{p}_{i}}%
-m\right) G_{n}\left( p_{1},...,p_{n};q_{1},...,q_{n}\right) \left( 
\overleftarrow{\stackrel{\wedge }{q}_{i}}-m\right) \psi ^{a,b}\left(
q_{i}\right) ,  \label{3.11}
\end{equation}
and use equation (\ref{3.10}) we get $n$-particle matrix element of $S$%
-matrix for physical states on mass-shell\footnote{%
For simplicity we do not write down symmetrized expressions for all matrix
elements due to identity of particles.} 
\begin{equation}
A\equiv \stackunder{i=1}{\stackrel{n}{\sqcap }}\overline{\psi _{\alpha _{i}}}%
^{a,b}\left( p_{i}\right) S^{\alpha _{1}...\alpha _{n};\beta _{1}...\beta
_{n}}\left( p_{1},...,p_{n};q_{1},...,q_{n}\right) \psi _{\beta
_{i}}^{a,b}\left( q_{i}\right) ,  \label{3.12}
\end{equation}
where: $p_{i}^{0}=\left( \left( {\bf p}_{i}\right) ^{2}+m^{2}\right) ^{1/2}$
and\ $q_{i}^{0}=\left( \left( {\bf q}_{i}\right) ^{2}+m^{2}\right) ^{1/2}$.

In Appendix, point 1, we show that all terms in $S\left( x-y\right) \sim
\delta ^{4}\left( x-y\right) \left( 1-\left( \beta _{0}\right) ^{2}\right) $
do not contribute to matrix elements on mass-shell .

The same situation arises if one considers matrix elements for physical
states with many photons {\bf in the presence} of DKP-particles.

Thus one goes to a very important conclusion: we can start from ${\cal L}$%
-for\-mu\-la\-tion of DKP-theory forgetting about constraints and \ to
consider instead equation (\ref{2.49}) the following expression for
generating functional: 
\begin{eqnarray}
{\cal Z}\left( I,\stackrel{\_}{I},{\cal J}_{\mu }\right) &=&{\cal Z}%
_{0}^{-1}\int {\cal D}\psi {\cal D}\overline{\psi }{\cal D}A_{\mu }\exp
\left\{ -i\int d^{4}x\left( \overline{\psi }\left( i\beta _{\mu }D^{\mu
}-m\right) \psi \right. \right.  \nonumber \\
&&\left. \left. -\frac{1}{4}F_{\mu \nu }F^{\mu \nu }-\frac{1}{2\alpha }%
\left( \partial _{\mu }A^{\mu }\right) ^{2}+\overline{\psi }I+\stackrel{\_}{I%
}\psi +{\cal J}_{\mu }A^{\mu }\right) \right\}  \label{3.13}
\end{eqnarray}
and to get for physical matrix element of $S$-matrix the correct expression.

Expressed in another way we can use equation (\ref{5.2}) for $S$-matrix with
relativistic invariant definition of T-product of operators $\psi \left(
x\right) $ and $\overline{\psi }\left( y\right) $ in interaction
representation \cite{Akhiezer, Kinoshita}: 
\begin{equation}
\left\langle 0\right| T\psi \left( x\right) \overline{\psi }\left( y\right)
\left| 0\right\rangle =\frac{i}{\left( 2\pi \right) ^{4}m}\int
d^{4}xe^{-ipx}d^{4}p\left\{ \frac{\stackrel{\wedge }{p}\left( \stackrel{%
\wedge }{p}+m\right) }{p^{2}-m^{2}+i\varepsilon }-1\right\}  \label{3.14}
\end{equation}

3) We prove equality between equation (\ref{3.11}) and corresponding matrix
element $S$-matrix in KG- theory.

Starting from definition (\ref{3.8}) for GF in DKP-theory one write equation
(\ref{3.11}) in the form (we omit non-essential multipliers) 
\begin{eqnarray}
&&A \equiv \stackunder{q_{0}\rightarrow \left( \left( {\bf q}\right)
^{2}+m^{2}\right) ^{1/2}}{\stackunder{p_{0}\rightarrow \left( \left( {\bf p}%
\right) ^{2}+m^{2}\right) ^{1/2}}{\lim }}\int \stackunder{i=1}{\stackrel{n}{%
\sqcap }}\int dx_{i}dy_{i}\overline{\psi }_{p}^{a}\left( x_{i}\right) \left(
i\beta _{\mu }\overrightarrow{\partial _{i}^{\mu }}-m\right)  \nonumber \\
&&\times \left\langle 0\right| T\psi \left( x_{1}\right) ...\psi \left(
x_{n}\right) \psi \left( y_{1}\right) ...\overline{\psi }\left( y_{n}\right)
\left| 0\right\rangle \left( i\beta _{\mu }\overleftarrow{\partial }%
_{j}^{\mu }-m\right) \psi _{q}^{b}\left( y_{j}\right) ,  \label{3.15}
\end{eqnarray}
where $\psi \left( x_{j}\right) $ $\overline{\psi }\left( y_{i}\right) $ are
Heisenberg operators; $\psi _{p}^{a}\left( x\right) $ and $\overline{\psi }%
_{q}^{b}\left( y\right) $ are defined by equation (\ref{3.6}).

Utilizing equations (\ref{2.3}) and (\ref{3.14}) for expression of operator $%
\psi \left( x\right) $ and $\overline{\psi }^{a}\left( p\right) $ in
component form, consider the following term: 
\begin{eqnarray}
&&\int e^{ipx}d^{4}x\overline{\psi }^{a}\left( p\right) \left( i\beta _{\mu
}\partial ^{\mu }-m\right) \psi \left( x\right) =  \nonumber \\
&=&\int e^{ipx}d^{4}x\left[ \left( -m+\frac{i}{m}p^{\mu }\partial _{\mu
}\right) \varphi \left( x\right) -\left( \partial ^{\mu }+ip^{\mu }\right)
\varphi _{\mu }\right]  \label{3.16}
\end{eqnarray}

As $ip_{\mu }e^{ipx}=\frac{\partial }{\partial x^{\mu }}e^{ipx}$ we can
rewrite equation (\ref{3.16}) in the form 
\begin{eqnarray}
\int e^{ipx}d^{4}x\overline{\psi }^{a}\left( p\right) \left( i\beta _{\mu
}\partial ^{\mu }-m\right) \psi \left( x\right) =\int e^{ipx}d^{4}x\frac{%
\left( -1\right) }{m}\left( \overrightarrow{\square }+m^{2}\right) \varphi
\left( x\right)  \nonumber \\
\hspace*{3cm}+\int d^{4}x\frac{\partial }{\partial x^{\mu }}\left[
e^{ipx}\left( \frac{\partial ^{\mu }}{m}\varphi -\varphi ^{\mu }\right) %
\right]  \label{3.17}
\end{eqnarray}

In the appendix one shows that the second term in equation \ (\ref{3.17})\
is equal to zero and thus physical matrix elements of scattering scalar
charged particles coincide in DKP and KG theories\footnote{%
The case of scattering charged particles by external EM field is a
particular one and the equivalence of the both theories follows the general
formula (\ref{A.5}) in Appendix.}.

Now we will prove the equality in both theories of many photons GF.

It is easy to show that generating functional of GF in KG theory has the
form: 
\begin{eqnarray}
{\cal Z}\left( {\cal J}^{\ast },{\cal J},{\cal J}_{\mu }\right) &=&{\cal Z}%
_{0}^{-1}\int {\cal D}A_{\mu }\exp \left\{ -i\int d^{4}x\text{Tr}\left( \ln 
\frac{G\left( x,x,A\right) }{G\left( x,x,0\right) }-\frac{1}{4}F_{\mu \nu
}F^{\mu \nu }\right. \right.  \nonumber \\
&&\left. -\frac{1}{2\alpha }\left( \partial _{\mu }A^{\mu }\right) ^{2}+%
{\cal J}_{\mu }A^{\mu }\right.  \nonumber \\
&&\left. \left. -\int d^{4}y{\cal J}^{\ast }\left( x\right) G\left(
x,y,A\right) {\cal J}\left( y\right) \right) \right\} ,  \label{3.18}
\end{eqnarray}
where GF $G\left( x,y,A\right) $ is defined in equation (\ref{2.35a}).

To get the generating functional of GF only for photons we have to put $%
{\cal J}^{\ast }={\cal J}=0$ in equation (\ref{3.18}) and $I=\stackrel{\_}{I}%
=0$ in equation (\ref{2.49}). \ Equality of these equations will be
established if we prove that 
\begin{equation}
{\cal Z}_{A}\equiv \det \frac{S\left( x,y,A\right) }{S\left( x,y,0\right) }%
=\exp \text{Tr}\ln \frac{S\left( x,x,A\right) }{S\left( x,x,0\right) }=\det 
\frac{G\left( x,y,A\right) }{G\left( x,y,0\right) }.  \label{3.19}
\end{equation}

On the other hand 
\begin{equation}
{\cal Z}_{A}={\cal Z}_{0}^{-1}\int {\cal D}\psi {\cal D}\overline{\psi }\exp
\left\{ i\int d^{4}x\overline{\psi }\left( i\stackrel{\wedge }{D}-m\right)
\psi \right\} ,  \label{3.20}
\end{equation}
where 
\begin{equation}
{\cal Z}_{0}=\int {\cal D}\psi {\cal D}\overline{\psi }\exp \left\{ i\int
d^{4}x\overline{\psi }\left( i\stackrel{\wedge }{\partial }-m\right) \psi
\right\} .  \label{3.20a}
\end{equation}

In component form expression (\ref{3.20}) equals to: 
\begin{eqnarray*}
{\cal Z}_{A} &=&{\cal Z}_{0}^{-1}\int {\cal D}\varphi {\cal D}\varphi ^{\ast
}{\cal D}\varphi _{\mu }{\cal D}\varphi _{\mu }^{\ast }\exp \left\{ i\int
d^{4}x(\varphi ^{\ast \mu }D_{\mu }\varphi -\varphi ^{\ast }D_{\mu }\varphi
^{\mu }\right. \\
&&\left. -m\left( \varphi ^{\ast }\varphi +\varphi ^{\ast \mu }\varphi _{\mu
}\right) )\right\} ,
\end{eqnarray*}

After integration over $\varphi _{\mu }^{\ast }$ and $\varphi ^{\mu }$ we
get 
\begin{equation}
{\cal Z}_{A}=\widetilde{{\cal Z}}_{0}^{-1}\int {\cal D}\varphi {\cal D}%
\varphi ^{\ast }\exp \left\{ -\frac{i}{m}\int d^{4}x\varphi ^{\ast }\left(
D^{\mu }D_{\mu }+m^{2}\right) \varphi \right\} ,  \label{3.21}
\end{equation}
where now 
\begin{equation}
\widetilde{{\cal Z}}_{0}^{-1}=\int {\cal D}\varphi {\cal D}\varphi ^{\ast
}\exp \left\{ -\frac{i}{m}\int d^{4}x\varphi ^{\ast }\left( \partial ^{\mu
}\partial _{\mu }+m^{2}\right) \varphi \right\} .  \label{3.21a}
\end{equation}

Doing substitution the $\frac{\varphi }{\sqrt{m}}\rightarrow \varphi $ we
see that the determinant (\ref{3.21}) is equal\ to the right hand side of
equation (\ref{3.19}).

The equivalence was proved.

\section{Reduction formulas in DKP formalism and e\-qui\-va\-len\-ce with
the KG Theory}

One writes down Yang-Feldman equations for Heisenberg operators in DKP
formalism. 
\begin{equation}
\psi \left( x\right) =\psi _{in}\left( x\right) +\int S_{R}\left( x-y\right)
j\left( y\right) d^{4}y  \label{4.1}
\end{equation}
\begin{equation}
\psi \left( x\right) =\psi _{out}\left( x\right) +\int S_{A}\left(
x-y\right) j\left( y\right) d^{4}y  \label{4.2}
\end{equation}

Here $\psi _{in}\left( x\right) $ and $\psi _{out}\left( x\right) $ satisfy
the free DKP equation: 
\begin{equation}
\left( i\beta _{\mu }\partial ^{\mu }-m\right) \psi _{in,out}\left( x\right)
=0  \label{4.3}
\end{equation}

$S_{R,A}$ are the retarded and advanced GF of the free equation 
\begin{equation}
\left( i\beta _{\mu }\partial ^{\mu }-m\right) S_{R,A}\left( x\right)
=\delta ^{4}\left( x\right) ,  \label{4.4}
\end{equation}
\begin{equation}
S_{R,A}\left( x\right) =\frac{1}{\left( 2\pi \right) ^{4}m}\int
e^{-ipx}\left( \frac{\stackrel{\wedge }{p}\left( \stackrel{\wedge }{p}%
+m\right) }{p^{2}-m^{2}\pm i\varepsilon p_{0}}-1\right) .  \label{4.5}
\end{equation}

$S_{R,A}$ is not equal to zero in up and down light cones. From equations (%
\ref{4.1}) to (\ref{4.5}) we obtain: 
\begin{equation}
\left( i\beta _{\mu }\partial ^{\mu }-m\right) \psi \left( x\right) =j\left(
x\right) .  \label{4.6}
\end{equation}

Choosing for $\psi _{in,out}$ ($\overline{\psi }_{in,out}$) two linearly
independent solutions (\ref{3.1}) and (\ref{3.2}) we can write (we omit
index $x$ and equation for $\overline{\psi }_{in,out}$) 
\begin{equation}
\psi _{in,out}=\frac{1}{\left( 2\pi \right) ^{3/2}}\int d^{3}p\left\{
a_{in,out}^{-}\left( {\bf p}\right) e^{-ipx}\psi ^{a}\left( {\bf p}\right)
+b_{in,out}^{+}\left( {\bf p}\right) e^{ipx}\psi ^{b}\left( {\bf p}\right)
\right\}  \label{4.7}
\end{equation}

Analagously we can use equations (\ref{3.6}) and (\ref{3.7}) to express $%
a_{in,out}^{\pm }\left( {\bf p}\right) $ and $b_{in,out}^{\pm }\left( {\bf p}%
\right) $ through the scalar product 
\begin{equation}
a_{in,out}^{-}\left( {\bf p}\right) =\int d^{3}x\psi ^{\ast a}\left(
x\right) \beta _{0}\psi _{in,out}\left( x\right) ,  \label{4.8}
\end{equation}
and so on. By definition 
\begin{equation}
a_{out}^{\pm }=S^{\dagger }a_{in}^{\pm }S.  \label{4.8a}
\end{equation}

We also can define Hisenberg operators of creation and annihilation 
\begin{equation}
a^{-}\left( {\bf p},x_{0}\right) =\int d^{3}x\psi ^{\ast a}\left( x\right)
\beta _{0}\psi \left( x\right) ,  \label{4.9}
\end{equation}
\begin{equation}
b^{+}\left( {\bf p},x_{0}\right) =\int d^{3}x\psi ^{\ast b}\left( x\right)
\beta _{0}\psi \left( x\right) ,  \label{4.10}
\end{equation}
and others; where $\psi \left( x\right) $ is the solution of equation (\ref
{4.6}).

Now it is possible to prove \cite{LSZ}, starting from equations (\ref{4.1}),
(\ref{4.2}), (\ref{4.9}) and (\ref{4.10}), that operators $a^{\pm }\left( 
{\bf p},x_{0}\right) $ and $b^{\pm }\left( {\bf p},x_{0}\right) $ have the
limits (in weak sense) for any matrix elements over total system physical
states, $\left| n_{in}\right\rangle $ $\left| m_{out}\right\rangle $, 
\begin{equation}
\stackunder{x_{0}\rightarrow \pm \infty }{\lim }\left\langle m_{out}\right|
a^{\pm }\left( {\bf p},x_{0}\right) \left| n_{in}\right\rangle =\left\langle
m_{out}\right| a_{out,in}^{\pm }\left| n_{in}\right\rangle ,  \label{4.11}
\end{equation}
and analogously for $b^{\pm }\left( {\bf p},x_{0}\right) $.

For simplicity we carry out the proof of equivalence for matrix elements of
scattering scalar particle (with positive charge from arbitrary initial
state $\left| n_{in}\right\rangle $ to final one $\left\langle
m_{out}\right| $. Using equation (\ref{4.11}) we have 
\begin{eqnarray}
&&\left\langle m_{out}\right| a_{out}^{-}\left( {\bf q}\right)
a_{in}^{+}\left( {\bf p}\right) \left| n_{in}\right\rangle =\left\langle
m_{out}\right| a_{in}^{-}Sa_{in}^{+}\left| n_{in}\right\rangle =  \nonumber
\\
&=&\stackunder{x_{0}\rightarrow -\infty }{\lim }\left\langle m_{out}\right|
a_{out}^{-}\left( {\bf q}\right) \int d\overrightarrow{x}\psi ^{\ast
a}\left( x\right) \beta _{0}\psi \left( x\right) \left| n_{in}\right\rangle 
\nonumber \\
&=&\left\langle m_{out}\right| a_{out}^{-}\left( {\bf q}\right) \left(
-i\right) \int d\overrightarrow{x}dx_{0}i\frac{\partial }{\partial x_{0}}%
\left( \overline{\psi }^{a}\left( x\right) \beta _{0}\psi \left( x\right)
\right) \left| n_{in}\right\rangle  \nonumber \\
&=&\frac{1}{i\left( 2\pi \right) ^{3/2}}\int d^{4}xe^{ipx}\overline{\psi }%
_{\alpha }^{a}\left( p\right) \left( i\beta _{\mu }\stackunder{x}{%
\overrightarrow{\partial }}^{\mu }-m\right) _{\alpha \beta }\left\langle
m_{out}\right| a_{out}^{-}\left( {\bf q}\right) \psi _{\beta }\left(
x\right) \left| n_{in}\right\rangle  \nonumber \\
&=&-\frac{1}{\left( 2\pi \right) ^{3}}\int d^{4}xd^{4}ye^{ipx}\overline{\psi 
}_{\alpha }^{a}\left( p\right) \left( i\beta _{\mu }\stackunder{x}{%
\overrightarrow{\partial }}^{\mu }-m\right) _{\alpha \beta }\left\langle
m_{out}\right| T\left( \psi _{\alpha }\left( x\right) \psi _{\beta }\left(
y\right) \right) \left| n_{in}\right\rangle  \nonumber \\
&\times &\left( i\beta _{\mu }\stackunder{y}{\overleftarrow{\partial }}^{\mu
}+m\right) _{\delta \gamma }\psi _{\gamma }^{b}\left( q\right)
e^{-ipy}\equiv \left\langle m_{out},{\bf q}\right| {\bf p},n_{in}\rangle
\label{4.12}
\end{eqnarray}

If we utilize the LSZ method for all particles states in $\left|
n_{in}\right\rangle $ and $\left\langle m_{out}\right| $ we get the equation
(\ref{3.15}).

Further, the proof of equivalence between DKP and KG goes in the same way as
in the end of Section 3 (see equations (\ref{3.17}) to (\ref{3.21}) and
Appendix).

\section{Conclusions}

1) Starting from canonical approach to DKP theory interacting with quantized
EM\ field and constructing the generating functional for GF of the theory we
strictly proved total equivalence between physical matrix elements of $S$%
-matrix in DKP and KG theories and between many photons GF in both theories.

The proof of equivalence between both theories have \ been carried out
utilizing the more general approach of Lehmann, Symanzik and Zimmermann by
reduction formalism.

We also proved the equivalence of the both theories, starting from
Lagrangian approach to generating functional in DKP theory(see equation (\ref
{3.13})) and forgetting about constraints.

2) In principle, the DKP as well as KG theories are nonrenormalizable ones
even for scalar particles due to the logarithmical divergence of one loop
diagrams of scattering two particles with exchange of two photons \cite
{Akhiezer}.

As it is well known that KG theory becomes renormalizable if we introduce a
self interaction term $\sim \lambda \left( \varphi ^{\ast }\varphi \right)
^{2}$. This problem can be solved in DKP theory in the same way: it is \
necessary to add to ${\cal L}$ in equation (\ref{2.1}) terms 
\begin{equation}
\lambda \left( \overline{\psi }P\psi \right) ^{2}=\lambda \left( \varphi
^{\ast }\varphi \right) ^{2},  \label{5.1}
\end{equation}
where $P=\stackunder{\mu }{\sqcap }\left( \beta _{\mu }\right) ^{2}$ is the
projector on the scalar part of $\psi $-function; $P$ is pseudoscalar .

3) In the framework the same method (sections 3,4) formally it is possible
to prove equivalence between DKP and Proca equation for spin one particles,
distructing from nonrenormalizability of these theories.

4)\ We would like to stress that DKP theory until now did not find wilder
application \ although this theory has some advantages just due to the
degenaration of $\beta _{\mu }$ matrices (one very simple to calculate trace
that of) and due to minimal character of interaction with EM\ fields. One
compares expressions for $S$-matrix in both theories: 
\begin{equation}
S_{DKP}=T\exp \left\{ i\int e\overline{\psi }\beta _{\mu }A^{\mu }\psi
d^{4}x\right\}  \label{5.2}
\end{equation}
\begin{equation}
S_{KG}=T\exp \left\{ i\int ie\left( \varphi ^{\ast }\partial _{\mu }\varphi
-\partial _{\mu }\varphi ^{\ast }\varphi -e\varphi ^{\ast }A_{\mu }\varphi
\right) A^{\mu }d^{4}x\right\}  \label{5.3}
\end{equation}

In the last case the interaction contains proportional \ terms to $e$\ and $%
e^{2\text{ }}$. Due to this in higher (two and more loops) approximations
combinatorial coefficients \ given to order $\ e^{2\text{ \ }}$before having
a complicated form.

\bigskip 5) About equivalence of DKP and KG for descripition of unstable
particles we would like to note that if we can apply conception of
asymptotic states to some such particle and utilize for physical matrix
elements of $S$-matrix the same method which have been used in Sections 3
and 4, then the proof of equivalence is obvious: for instance, for decay of $%
K_{l}$ $-mesons$ we can calculate the imaginary part of the GF of $K_{l}-$ $%
meson$ and get equivalence with exactness redefining the $\varphi $
component of DKP $\psi $ function: $i\varphi _{KG}=\varphi _{DKP}/m$, see
equation (\ref{3.17}).

\section{Appendix}

\setcounter{footnote}{0}

1) One proves that all terms $\sim \left( 1-\left( \beta _{0}\right)
^{2}\right) \delta ^{4}\left( x-y\right) $ in GF of scalar particles,
equation (\ref{2.49}), do not contribute to physical matrix elements (see
phrase after equation (\ref{3.12})). First, consider the simplest case: GF
in external EM field $A_{\mu }$, which is equal 
\begin{equation}
i\left\langle 0\right| T\psi \left( x\right) \overline{\psi }\left( y\right)
\left| 0\right\rangle =S\left( x,y,A\right) +\frac{1}{m}\left( 1-\left(
\beta _{0}\right) ^{2}\right) \delta ^{4}\left( x-y\right) ,  \label{A.1}
\end{equation}
where $S\left( x,y,A\right) =\left( i\beta _{\mu }D^{\mu }-m\right)
^{-1}\delta ^{4}\left( x-y\right) $.

By definition, matrix elements of scattering particles with positive charge
by external EM field is: 
\begin{eqnarray}
&&\stackunder{q_{0}\rightarrow \sqrt{\left( {\bf q}\right) ^{2}+m^{2}}}{%
\stackunder{p_{0}\rightarrow \sqrt{\left( {\bf p}\right) ^{2}+m^{2}}}{\lim }}%
\frac{\left( -1\right) }{\left( 2\pi \right) ^{3}}\int d^{4}xd^{4}y\overline{%
\psi }_{p}^{a}\left( x\right) \left( i\beta _{\mu }\stackunder{x}{%
\overrightarrow{\partial }}^{\mu }-m\right)  \nonumber \\
&\times &\left\langle 0\right| \psi _{\alpha }\left( x\right) \overline{\psi 
}_{\beta }\left( y\right) \left| 0\right\rangle \left( i\beta _{\mu }%
\stackunder{y}{\overleftarrow{\partial }}^{\mu }+m\right) \psi
_{q}^{b}\left( y\right)  \label{A.2}
\end{eqnarray}

Inserting equation (\ref{A.1}) in equation (\ref{A.2}) we see that the term $%
\sim \delta ^{4}\left( x-y\right) $ does not contribute to result since, by
definition, $\delta -$ function we must change directions of arrows on the
inverse and get zero: 
\begin{equation}
\overline{\psi }_{p}^{a}\left( x\right) \left( i\beta _{\mu }\stackunder{x}{%
\overleftarrow{\partial }}^{\mu }+m\right) =0,\ \left( i\beta _{\mu }%
\stackunder{y}{\overrightarrow{\partial }}^{\mu }-m\right) \psi
_{q}^{b}\left( y\right) =0  \label{A.3}
\end{equation}

In the case of quantized EM field we have to start from equation (\ref{2.49}%
). So far as external photons do not influence on the appearance of terms $%
\sim \delta ^{4}\left( x-y\right) $ it is enough to consider the case ${\cal %
J}_{\mu }=0$\footnote{%
This case is considered in equations (\ref{3.18}) to (\ref{3.21a}).} in
equation (\ref{A.1}).

Thus any GF of scalar particles expressed through symmetrized product of one
particle GF (\ref{A.1}) under integral over $A_{\mu }$ in equation (\ref
{3.18}) 
\begin{eqnarray}
&&\left\langle 0\right| T\psi \left( x_{1}\right) ...\psi \left(
x_{n}\right) \overline{\psi }\left( y_{1}\right) ...\overline{\psi }\left(
y_{n}\right) \left| 0\right\rangle =  \nonumber \\
&=&{\cal Z}_{0}^{-1}\int {\cal D}A_{\mu }\exp \left\{ -i\int d^{4}x\left(
\det \ln \frac{S\left( x,x,A\right) }{S\left( x,x,0\right) }\right. \right. 
\nonumber \\
&&\left. \left. -\frac{1}{4}F_{\mu \nu }F^{\mu \nu }-\frac{1}{2\alpha }%
\left( \partial _{\mu }A^{\mu }\right) ^{2}\right) \right\}  \nonumber \\
&&\times \left( \sum\limits_{p_{i,j}}\stackunder{i,j=1}{\stackrel{n}{\sqcap }%
}\left( S\left( x_{i},y_{j},A\right) -\frac{1}{m}\left( 1-\left( \beta
_{0}\right) ^{2}\right) \delta ^{4}\left( x_{i}-y_{j}\right) \right) \right)
,
\end{eqnarray}
where $\sum\limits_{p_{i,j}}$ means summation over all permutations $x_{i}$
or $y_{j}$.

All $\delta ^{4}\left( x_{i}-y_{j}\right) $ we can wear out from sign of $%
\int {\cal D}A_{\mu }$ .

After transition to matrix elements of $S$-matrix we can apply to these
terms the same procedure which we used for equation (\ref{A.2}) and get zero.

Thus we proved that terms $\sim \left( 1-\left( \beta _{0}\right)
^{2}\right) \delta ^{4}\left( x_{i}-y_{j}\right) $ do not contribute to
physical matrix elements of $S$-matrix and do not violate relativistic
invariance and microcausality for physical observable values.

2) Proof that the last term under total derivative in equation (\ref{3.17})
which can contain quasilocal one is equal zero.

One writes down all terms for matrix element (\ref{4.12}) in {\bf component}
form, utilizing equation (\ref{3.17}) we get\footnote{%
No essential multipliers are omitted} 
\begin{eqnarray}
\left\langle m_{out},{\bf q}\right| \left. {\bf p},n_{in}\right\rangle
&\cong &\int d^{4}xd^{4}y\left\langle m_{out}\right| \exp \left\{ i\left(
px-qy\right) \right\} \left( \square _{x}+m^{2}\right) \left( \square
_{y}+m^{2}\right)  \nonumber \\
&&\times T\left( \varphi \left( x\right) \varphi ^{\ast }\left( y\right)
\right) -\stackunder{x}{\partial }_{\mu }\left[ e^{ipx}\left( \square
_{y}+m^{2}\right) \right.  \nonumber \\
&&\left. \left( \frac{\stackunder{x}{\partial }^{\mu }}{m}T\left( \varphi
\left( x\right) \varphi \left( y\right) \right) -T\left( \varphi ^{\mu
}\left( x\right) \varphi ^{\ast }\left( y\right) \right) \right) \right]
e^{-iqy}  \nonumber \\
&&+\stackunder{y}{\partial }_{\mu }\left[ e^{-iqy}\left( \square
_{x}+m^{2}\right) \right.  \nonumber \\
&&\left. \left( \frac{\stackunder{y}{\partial }^{\mu }}{m}T\left( \varphi
\left( x\right) \varphi ^{\ast }\left( y\right) \right) -T\left( \varphi
\left( x\right) \varphi ^{\ast \mu }\left( y\right) \right) \right) \right]
e^{ipx}  \nonumber \\
&&+\stackunder{x}{\partial }_{\mu }\stackunder{y}{\partial }_{\nu }\left[
e^{ipx-iqy}\left( T\left( \varphi ^{\mu }\left( x\right) \varphi ^{\ast \nu
}\left( y\right) \right) \right. \right.  \nonumber \\
&&\left. -\frac{\stackunder{x}{\partial }^{\mu }\stackunder{y}{\partial }%
^{\nu }}{m^{2}}T\left( \varphi \left( x\right) \varphi ^{\ast }\left(
y\right) \right) -\frac{\stackunder{x}{\partial }^{\mu }}{m}T\left( \varphi
\left( x\right) \varphi ^{\ast \nu }\left( y\right) \right) \right. 
\nonumber \\
&&\left. \left. \frac{\stackunder{y}{\partial }^{\nu }}{m}T\left( \varphi
^{\mu }\left( x\right) \varphi ^{\ast \nu }\left( y\right) \right) \right) 
\right] \left| \overrightarrow{p},n_{in}\right\rangle  \label{A.5}
\end{eqnarray}

The quasilocal term containing $\delta ^{4}\left( x-y\right) $ arises only
from the seventh one: 
\[
\stackunder{x}{\partial }^{\mu }\stackunder{y}{\partial }^{\nu }T\left(
\varphi \left( x\right) \varphi ^{\ast }\left( y\right) \right) =T\left(
\partial ^{\mu }\varphi \left( x\right) \partial ^{\nu }\varphi ^{\ast
}\left( y\right) \right) -ig^{0\mu }g^{0\nu }\delta ^{4}\left( x-y\right) . 
\]

Here we used that $\delta ^{4}\left( x_{0}-y_{0}\right) \left[ \varphi
\left( x\right) ,\varphi ^{\ast }\left( y\right) \right] =-i\delta
^{4}\left( x-y\right) $. Thus, 
\begin{eqnarray*}
&&\int d^{4}xd^{4}y\stackunder{x}{\partial }^{\mu }\stackunder{y}{\partial }%
^{\nu }\left( e^{ipx-iqy}\frac{\stackunder{x}{\partial }^{\mu }\stackunder{y%
}{\partial }^{\nu }}{m^{2}}T\left( \varphi \left( x\right) \varphi ^{\ast
}\left( y\right) \right) \right) = \\
&=&-\frac{i}{m^{2}}\left( 2\pi \right) ^{3}\delta \left( {\bf p}-{\bf q}%
\right) \int dx_{0}dy_{0}\stackunder{x}{\partial }_{0}\stackunder{y}{%
\partial }_{0}\left( \exp \left( ip_{0}x_{0}-iq_{0}y_{0}\right) \delta
\left( x_{0}-y_{0}\right) \right) =0.
\end{eqnarray*}

\section{Acknowledgments}

We would like to thank I. V. Tyutin, who suggested to use reduction
formalism of LSZ (see Section 4) to proof the equivalence of both theories,
and D. M. Gitman for stimulating criticism .

V.Ya.F. thanks to FAPESP for support (grant number 98/06237-0) and RFFI for
partial support (grant number 99-01-00376). B. M. P. thanks to CNPq for
partial support.


\begin{thebibliography}{99}
\bibitem{Petiau}  G. Petiau, Acad. Roy. de Belg., A. Sci. Mem. Collect {\bf %
16} (1936).

\bibitem{Duffin}  R.Y. Duffin, Phys.Rev. {\bf 54 }(1938) 1114.

\bibitem{Kemmer}  N. Kemmer. Proc. Roy. Soc {\bf A173 }(1939) 91.

\bibitem{Umezawa}  H. Umezawa, {\it Quantum Field Theory},
North-Holland,1956.

\bibitem{Akhiezer}  A.I. Akhiezer and V.B. Berestetski, {\it Quantum
Electrodynamics} 2nd ed. Inter.,New York, 1965.

\bibitem{Kinoshita}  T. Kinoshita, Prog. Theor. Phys. {\bf 5 }(1950) 473 ;
T. Kinoshita and Y. Nambu, ibid.{\bf \ 5 }(1950) 749.

\bibitem{pimentel}  B.M. Pimentel and J.L. Tomazelli, Prog. Theor. Phys. 
{\bf 45} (1995) 1105.

\bibitem{Nieto}  R.A. Krajcik and M.M. Nieto, Am. Journ. of Physics, {\bf 45 
}(1974) 818.

\bibitem{Ginzburg}  V.L. Ginzburg, {\it Quantum Field Theory and Quantum
Statistics}, Essays in Honor of the Siextieth Birthday of E.S. Fradkin, {\bf %
Vol. 2} (1987) 15, IOP Publishing Ltd. 1987, Adam Hilger, Bristol.

\bibitem{Wightman}  A. Wightman, {\it Aspects of Quantum Theory}, p.95,
Edited by A. Salam and E.P. Wigner, Cambridge, University Press (1972).

\bibitem{Velo}  G. Velo and D. Zwanziger, Phys. Rev. {\bf 186 }(1969) 1337 ;
ibid. {\bf 188 }(1969) 2218.

\bibitem{Niet2}  E. Fischbach, et al., Phys. Rev. Lett. {\bf 26 }(1971)1200.

\bibitem{LSZ}  H. Lehmann, K. Symanzik and W. Zimmermann, Nuovo Cim. {\bf 1 }%
(1955) 425.

\bibitem{Gitman}  D.M. Gitman and I.V. Tyutin, {\it Quantization of Fields
with Constraints}, Springer-Verlag, New York/Berlin, 1990

\bibitem{Dirac}  P.A.M. Dirac, Proc. Roy. Soc. {\bf A246} (1958) 326.

\bibitem{Fradkin}  So called Batalin-Fradkin-Vilkovisky method , see
references on the method in book \cite{Gitman}.

\bibitem{Faddeev}  L. Faddeev and A. Slanov, {\it Gauge Fields, Introduction
to Quantum Theory}, The Benjamin Cummings Publishing Company, 1980.
\end{thebibliography}
\end{document}